\documentclass[aps,pra,twocolumn,a4paper,groupedaddress,floatfix,10pt]{revtex4-1}
\usepackage{graphicx,amsmath,amssymb,amsfonts,dsfont,enumitem}

\newcommand{\mc}[1]{\ensuremath{\mathcal{#1}}}

\usepackage{color}

\newcommand{\rc}{r_{\text{C}}}
\newcommand{\mm}{\phantom{$-$}}

\begin{document}

\title{A polynomial Ansatz for Norm-conserving Pseudopotentials}

\author{Martin Kiffner${}^{1,2}$}
\author{Dieter Jaksch${}^{2,1}$}
\author{Davide Ceresoli${}^{3}$}

\affiliation{Centre for Quantum Technologies, National University of Singapore,
3 Science Drive 2, Singapore 117543${}^1$}
\affiliation{Clarendon Laboratory, University of Oxford, Parks Road, Oxford OX1
3PU, United Kingdom${}^2$}
\affiliation{CNR-ISTM Istituto di Scienze e Tecnologie Molecolari and INSTM UdR di Milano, via Golgi 19,
20133 Milano, Italy${}^3$}

\begin{abstract}
We show that efficient norm-conserving pseudopotentials for electronic structure calculations can be obtained from a  polynomial Ansatz for the 
potential. Our pseudopotential is a polynomial of degree ten in the radial variable and fulfils  the same smoothness 
conditions  imposed by the Troullier-Martins method [Phys. Rev. B \textbf{43}, 1993 (1991)] where pseudopotentials are represented by a 
polynomial of degree twenty-two. 
We compare our method to the Troullier-Martins approach  in electronic structure calculations for diamond and iron in the $\text{bcc}$ structure 
and find that the two methods perform equally well in calculations of the total energy. However,  
first and second derivatives of the total energy with respect to atomic coordinates converge significantly faster with the plane 
wave cutoff if the standard Troullier-Martins potentials are replaced by the pseudopotentials introduced here.
\end{abstract}

\maketitle

\section{Introduction \label{intro}}
Density Functional Theory (DFT)~\cite{hohenbergkohn:64,kohnsham:65} 
is nowadays the most common computational method for calculating electronic 
structure properties of molecules and solids from first principles. 
The standard formulation of DFT is based on local, semilocal and 
hybrid exchange-correlation functionals~\cite{martin:es}, and has proven to be very successful
in describing the electronic structure of weakly-correlated systems~\cite{Kohn1999,Pople1999,Bowler2016,Lejaeghereaad2016}. 

A second pillar of  DFT calculations is the concept of pseudopotentials~\cite{martin:es}.  
They allow one to reduce the numerical cost of electronic structure calculations significantly by eliminating atomic core states 
that do not participate  in chemical bonding. 
Modern DFT calculations are based on norm-conserving 
pseudopotentials (NCPPs)~\cite{hamann:79,bachelet:82,vanderbilt:85,kerker:80,troullier:91,rappe:90,reis:03}, 
ultrasoft pseudopotentials (USPPs)~\cite{Vanderbilt1990,Laasonen1993,Kresse1994,Moroni1997,Garrity2014} or
projector augmented wave (PAW) methods~\cite{Blochl1994,Holzwarth1998,Kresse1999}.

 Calculations based on NCPPs are typically 
carried out in the Kleinman-Bylander approach~\cite{kleinman:82} where a set of pseudopotentials for 
different angular momenta are  represented by one local potential and a set of separable, nonlocal projectors. 
This method enables computationally efficient electronic structure calculations based on simple plane-wave 
representations of the Hamiltonian~\cite{espresso}. 
The USPP and PAW methods have been developed to further improve  (i) the transferability of
pseudopotentials among different chemical environments
and oxidation states and (ii) the numerical efficiency of
ab initio calculations.
However, NCPPs are significantly easier to implement than USPPs and PAWs 
due to the reconstruction/augmentation term in these methods~\cite{Miwa2011}.
Hence NCPPs remain the method of choice in more advanced calculations like, e.g.,  density-functional perturbation 
theory~\cite{baroni:01} or many-body perturbation theory~\cite{hedin:65,hybertsen:85}, and 
further improving their efficiency is a highly desirable goal~\cite{hamann:13}. 
One of the most successful methods for generating NCPPs is the Troullier-Martins (TM) method~\cite{troullier:91}. 
At the heart of this approach is an Ansatz for the pseudowavefunction (PWF), and its parameters 
are chosen such that norm-conservation and a set of smoothness conditions are met~\cite{troullier:91}. 
Like in other NCPP methods~\cite{hamann:79,bachelet:82,vanderbilt:85,kerker:80,rappe:90}, 
the pseudopotential itself is obtained from the PWF by an inversion of the radial 
Schr\"odinger equation. In the case of the TM method, the pseudopotential turns out to be 
a polynomial of degree twenty-two in the radial variable due to the specific Ansatz for the PWF. 

Here we pursue a different approach and show that efficient NCPPs can be obtained from a polynomial Ansatz for the pseudopotential. 
In this way, the final step of inverting  the radial Schr\"odinger equation is not required. 
Our pseudopotentials are represented by a polynomial of degree ten  in the radial variable and we impose 
the same smoothness conditions as in the TM method. 

We carry out electronic structure calculations for diamond and iron with the  Quantum Espresso code~\cite{espresso} and 
analyse the performance of our pseudopotentials compared to the TM potentials. While both pseudopotential methods 
perform equally well in calculations of the total energy and pressure, 
we find that our pseudopotentials speed up the convergence of calculations for atomic forces and phonon frequencies. 
Our pseudopotential scheme thus promises big computational savings when calculating structural relaxations and phonon frequencies 
in large systems.

Note that our pseudopotential method is related to unpublished work by von Barth and Car who suggested the generation of NCPPs by 
varying the parameters in an Ansatz for the pseudopotential. Their proposed parametrisation of the pseudopotential is different from ours 
and has been used, e.g., in~\cite{dalcorso:93}. 

This paper is organised as follows. In the Methods section~\ref{model}  we briefly review the theory of NCPPs (Sec.~\ref{general}) 
and the TM  method (Sec.~\ref{tm}). These two sections set the stage for  our  novel approach for constructing NCPPs which is described in Sec.~\ref{PA}. 
We then compare the performance of our NCPPs with the standard TM approach in electronic structure calculations for diamond and iron. 
These results are presented in Sec.~\ref{results}, and a brief summary  is given in Sec.~\ref{sum}. 
\section{Methods \label{model}}
In Sec.~\ref{general} we briefly review the general theory of NCPPs. The 
TM method~\cite{troullier:91} and our novel approach for generating NCPPs are described 
in Secs.~\ref{tm} and~\ref{PA}, respectively. 
\subsection{Norm-conserving pseudopotentials\label{general}}
The generation of pseudopotentials starts with an atomic DFT calculation 
which results in the self-consistent all-electron (AE) potential 
\begin{align}
V_{\text{AE}}[\rho](r) = V_{\text{C}}(r) +V_{\text{HT}}[\rho](r) + V_{\text{XC}}[\rho](r) \,,
\end{align}
where $V_{\text{C}}$ is the the Coulomb potential of the ion core, $V_{\text{HT}}$ is the Hartree energy 
of all electrons, $V_{\text{XC}}$ is the exchange-correlation potential and $\rho$ is the  electron density~\cite{martin:es}. 
For  non-relativistic calculations 
the AE radial wavefunctions $R_{nl}$ with principal quantum number $n$, orbital angular momentum $l$ and energy 
$E_{nl}$ are determined by the radial Schr\"odinger equation, 
\begin{align}
 \left[-\frac{\text{d}^2}{\text{d}r^2} +\frac{l(l+1)}{r^2} + V_{\text{AE}}(r) -E_{nl} \right]R_{nl}(r) r =0\,, 
\label{radialAE}
\end{align}
where $r$ is the radial coordinate of the electron. 
We introduce a short-hand notation for states $nl$ corresponding to valence states and denote their 
energies and radial wavefunctions by $E_{l}^{\text{(v)}}$ and  $R_l^{\text{(v)}}$, respectively. 
For each of  these valence states 
one introduces an  $l$-dependent pseudopotential $\mc{V}_l$ (we denote all pseudopotential quantities by caligraphic letters), and the  
nodeless  pseudo-wavefunction (PWF) $\mc{R}_{l}$ is the solution to the non-relativistic radial Schr\"odinger 
equation with potential $\mc{V}_l$ and energy $\mc{E}_l$, 
\begin{align}
 \left[-\frac{\text{d}^2}{\text{d}r^2} +\frac{l(l+1)}{r^2} + \mc{V}_l(r) -\mc{E}_{l} \right]\mc{R}_{l}(r) r =0. 
\label{radialPP}
\end{align}
In order to replace the AE potential by the set of $l$-dependent pseudopotentials in electronic structure calculations, 
the following conditions have to be met~\cite{hamann:79,martin:es}:
%
%
\begin{enumerate}[wide, labelwidth=!, labelindent=0pt]
\item[(NC1)] 
The ground state energy $\mc{E}_l$ of the pseudopotential has to coincide with the AE valence energy $E_{l}^{\text{(v)}}$, 
\begin{align}
 \mc{E}_l = E_{l}^{\text{(v)}}\,.
 \label{eval}
\end{align}
 \item[(NC2)] 
 The PWF must coincide with the AE wavefunction outside a  cutoff region $\rc$, 
\begin{align}
 \mc{R}_l(r)=R_l^{\text{(v)}}(r)\quad \text{for}\quad r\ge\rc\,, 
 \label{PWF}
\end{align}
and we have suppressed the $l$-dependence of  $\rc$ for simplicity. 
\item[(NC3)]
The norm of the PWF has to equal the norm of the AE wavefunction inside the core region, 
 \begin{align}
  \int\limits_0^{\rc} r^{2} |\mc{R}_l|^2  \text{d}r =\int\limits_0^{\rc} r^{2} |R_l^{\text{(v)}}|^2\text{d}r \,. 
  \label{nc}
 \end{align}
This norm-conservation condition  ensures that the first energy derivative of the logarithmic derivatives of the AE and pseudo 
wavefunctions agrees at $\rc$~\cite{hamann:79}. 
\end{enumerate}
Since  $\mc{R}_l$ is nodeless by construction, the pseudopotential $\mc{V}_l$ can be obtained from the PWF by inverting the radial Schr\"odinger equation~(\ref{radialPP}), 
\begin{align}
  \mc{V}_l(r) =   \mc{E}_{l} - \frac{l(l+1)}{r^2} + \frac{1}{\mc{R}_{l}(r) r}\frac{\text{d}^2}{\text{d}r^2} \left[\mc{R}_{l}(r) r\right] \,.
 \label{inverted}
\end{align}
Conditions (NC1) and (NC2) thus enforce that  $\mc{V}_l$  coincides with $V_{\text{AE}}$ outside the  cutoff region $\rc$. 
Conversely, $\mc{V}_l(r)=V_{\text{AE}}(r)$ for $r\ge \rc$ implies together with conditions (NC1), (NC3) and the normalisation of the radial wavefunction that 
condition (NC2) is met. An equivalent set of conditions is thus given by (NC1), (NC3) and   (NC2'), where (NC2')is defined as follows: 
 \begin{enumerate}[wide, labelwidth=!, labelindent=0pt]
\item[(NC2')] $\mc{V}_l$ must coincide with $V_{\text{AE}}$ outside the cutoff region, 
\begin{align}
 \mc{V}_l(r) =  \left\{ 
\begin{array}{l}
 V_{\text{AE}}(r) \quad \text{if}\quad r\ge \rc\,, \\[0.3cm]
  C_l(r) \quad \text{if} \quad r < \rc\,,
\end{array}
 \right.
 \label{PP}
\end{align}
where $C_l(r)$ is the potential inside the core. 
\end{enumerate}
In order to use the pseudopotential in electronic structure calculations one has to generate ionic pseudopotentials, 
\begin{align}
 \mc{V}_l^{\text{ion}}(r) = \mc{V}_l(r) - \mc{V}_{\text{HT}}(r) - \mc{V}_{\text{XC}}(r)\,,
\end{align}
where $\mc{V}_{\text{HT}}$ and $\mc{V}_{\text{XC}}(r)$ are the Hartree and exchange-correlation potentials calculated from the PWFs, respectively. 
The semilocal pseudopotential operator is then given by
\begin{align}
 \hat{\mc{V}}(r) = \sum\limits_l \mc{V}_l^{\text{ion}}(r) \hat{P}_l\,,
\end{align}
where $\hat{P}_l$ projects the wavefunction onto the $l$th angular momentum component. 
\subsection{Review of the Troullier-Martins method \label{tm}}
The TM method~\cite{troullier:91} for generating NCPPs makes the following Ansatz for the PWF, 
\begin{align}
 \mc{R}_l^{\text{TM}} = & 
 \left\{ 
\begin{array}{l}
 R_l^{\text{(v)}}(r) \quad \text{if}\quad r\ge \rc\,, \\[0.3cm]
 r^l \exp[p(r)] \quad \text{if} \quad r < \rc\,.
\end{array}
 \right.
 \label{Ansatz}
\end{align}
Outside the core region $\mc{R}_l^{\text{TM}}$ coincides with the AE wavefunction $R_l^{\text{(v)}}$, and hence 
(NC2) is automatically fulfilled. Inside the core region $\mc{R}_l^{\text{TM}}=r^l \exp[p(r)]$, where 
$p(r)$ is a polynomial of degree twelve  which contains only even powers of  $r$, 
\begin{align}
 p(r) = \sum\limits_{i=0}^{6} c_{2 i} r^{2 i}\,.
 \label{poly}
\end{align}
The seven coefficients $c_{2i}$ in Eq.~(\ref{poly}) are chosen such that the PWF and the potential exhibit the following smoothness conditions at $r=0$ and $r=\rc$: 
\begin{enumerate}[wide, labelwidth=!, labelindent=0pt]
 \item[(S1)] 
$\mc{V}_l^{\text{TM}}$ should have zero curvature  at the origin, i.e., $[\mc{V}_l^{\text{TM}}]''(0)=0$, where the double prime denotes the second derivative with respect to $r$.  
This condition can be cast into a non-linear equation for the coefficients $c_{2}$ and $c_4$ and reads~\cite{troullier:91}, 
\begin{align}
 c_2^2 +c_4(2l +5) =0 \,. 
\end{align}
\item[(S2)] 
$ R_l^{\text{TM}}$ and its first four derivatives should  be continuous at $\rc$. These five conditions   result 
in a system of linear equations for five of the coefficients $c_{2i}$ and can be solved with standard methods. 
\end{enumerate}
(S1) and (S2) constrain six out of the seven coefficients $c_{2i}$ in Eq.~(\ref{poly}). The last free parameter  is determined by condition (NC3) in Sec.~\ref{general}, 
 \begin{align}
  \int\limits_0^{\rc} r^{2(l+1)} \exp[2 p(r)]\text{d}r =\int\limits_0^{\rc} r^{2} |R_l^{\text{(v)}}|^2\text{d}r \,. 
  \label{nctm}
 \end{align}
This is a highly non-linear equation, and a solution is not guaranteed for all cutoff parameters $\rc$. 
The above procedure fully determines the PWF. 
Finally, $\mc{V}_l^{\text{TM}}$  is obtained from the PWF via Eq.~(\ref{inverted}) by setting $\mc{E}_l=E_{l}^{\text{(v)}}$ such that condition (NC1) is fulfilled. 
With the help  of Eq.~(\ref{Ansatz}), the pseudopotential  in the core region can be written as  
\begin{align}
  C_l^{\text{TM}}(r) = \mc{E}_l+2 \frac{l+1}{r}p'(r)+ p''(r)+[p'(r)]^2\,.
\end{align}
The definition of  $p$ in Eq.~(\ref{poly}) allows us to write $C_l^{\text{TM}}(r)$ as a polynomial in $r$, 
\begin{align}
   C_l^{\text{TM}}(r) = \sum\limits_{i=0}^{11} \mathcal{C}_{2 i} r^{2 i} \,.
 \label{C_TM}
\end{align}
We thus find that the pseudopotential in the core region is a polynomial of degree twenty-two, and $C_l^{\text{TM}}(r)$  contains only even powers of $r$. 
The explicit expressions for the coefficients $\mc{C}_{2i}$ in terms of $c_{2i}$ are given in Appendix~\ref{coeff}. 
\subsection{Variation of the potential \label{PA}}
Here we introduce a novel approach for generating NCPPs via the direct variation of a polynomial Ansatz (PA) for the potential. 
We label this potential by $\mc{V}_l^{\text{PA}}$ and associated quantities with the 
superscript PA. The pseudopotential outside the core region must  coincide with 
the AE potential such that condition (NC2') is satisfied. Inside the core region we make the following  Ansatz, 
\begin{align}
  C_l^{\text{PA}}(r) =  \sum\limits_{i=0}^{5} \mc{X}_{2 i} r^{2 i}\,.
 \label{C_PA}
\end{align}
$C_l^{\text{PA}}$ in Eq.~(\ref{C_PA}) contains only even powers of $r$  like $C_l^{\text{TM}}$ of the TM method in Eq.~(\ref{C_TM}). 
However,  our Ansatz for $C_l^{\text{PA}}$  is not equivalent to the TM method since the degree of the polynomial $C_l^{\text{PA}}$ is only ten 
and thus much lower than the degree of  $C_l^{\text{TM}}$. 

Next we show that our Ansatz for  $C_l^{\text{PA}}$ is sufficient for  creating a NCPP with the same smoothness conditions 
(S1) and (S2) of the TM method. First, condition (S1) can be satisfied by setting $\mc{X}_2 =0$ in Eq.~(\ref{C_PA}). The Ansatz 
in Eq.~(\ref{C_PA}) then contains only a constant term and polynomials that are at least of order 4, and hence the second derivative of 
$C_l^{\text{PA}}$ vanishes at $r=0$. 

Second, the five conditions in (S2)  concerning the continuity of the PWF and its derivatives at $\rc$ can be cast into conditions on the potential via 
Eq.~(\ref{inverted}). We find that $\mc{V}_l^{\text{PA}}$ and its first two derivatives must be continuous at $\rc$, and 
these three conditions result in a linear system of equations for three of the coefficients $\mc{X}_{2i}$. Here we choose 
to express $\mc{X}_6$, $\mc{X}_8$ and $\mc{X}_{10}$ in terms of $\mc{X}_{0}$ and $\mc{X}_{4}$, and  the explicit expressions for these coefficients are given in 
Appendix~\ref{coeff}.

By imposing the smoothness conditions (S1) and (S2) as described above, $C_l^{\text{PA}}$ only depends on the  two coefficients $\mc{X}_0$ and $\mc{X}_4$. 
These parameters must be chosen such that conditions (NC1) and (NC3) are satisfied. 
To this end we find the ground state energy and wavefunction of $\mc{V}_l^{\text{PA}}$ by solving the eigenvalue problem in Eq.~(\ref{radialPP}) 
on a discrete  grid. The optimal parameters $\mc{X}_0$ and $\mc{X}_4$ for satisfying conditions (NC1) and (NC3) 
are found by standard Newton methods, and we find fast convergence for suitable starting values. 
In particular, we find that the eigenvalues of our PA pseudopotentials match those of the AE calculation very accurately. 
For all systems considered in Sec.~\ref{results}, the maximal difference between  PA and AE eigenvalues 
is 0.5 meV for s orbitals, and for p and d orbitals the deviation is even smaller by at least one order of magnitude.
\section{Results \label{results}}
In order to test the   method introduced in Sec.~\ref{PA}, 
we compare the performance of the PA and TM pseudopotentials in electronic structure calculations for Carbon and Iron. 
All DFT calculations based on pseudopotentials are performed with the Quantum Espresso code~\cite{espresso} using the PZ functional~\cite{perdew:81}, and AE calculations are 
carried out with the same exchange-correlation functional and the ELK code~\cite{elk}.
%
%
\begin{figure}[t!]
\begin{center}
\includegraphics[width=8cm]{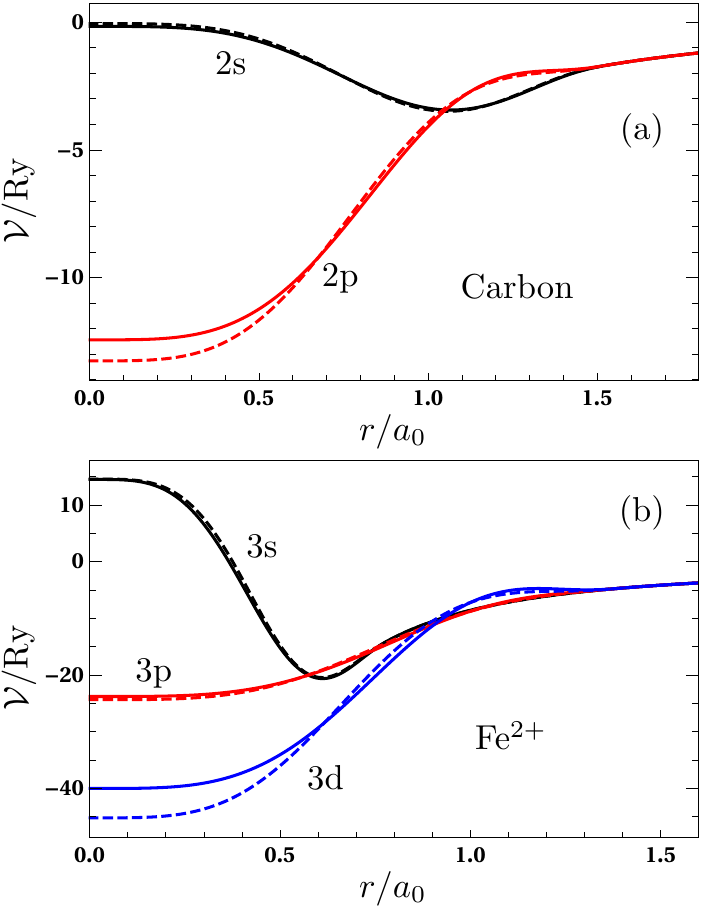}
\end{center}
\caption{\label{fig1}
(Color online) 
Comparison of the screened pseudopotentials generated by the PA and TM methods for (a) C and (b) Fe${}^{2+}$. 
The potentials of the PA method for s, p and d states are shown by black, red and blue solid lines, respectively,  
and the corresponding TM potentials  are shown by dashed lines. 
In (a),  the cutoff parameter is $\rc=1.54 a_0$ for all potentials, and $a_0$ is the Bohr radius. 
In (b), we set $\rc(l=0)= 0.8 a_0$ and $\rc(l=1)= \rc(l=2)= 1.4 a_0$.
}
\end{figure}
%

In a first step, we compare the screened pseudopotentials for the TM and PA methods. To this end we perform non-relativistic (semi-relativistic)  DFT calculations 
for a single C atom (Fe${}^{2+}$ ion). We then generate the TM and PA pseudopotentials for the valence states as described in Sec.~\ref{model}, and the results 
are shown in Fig.~\ref{fig1}. Note that we choose the same cutoff parameters for the TM and PA potentials for a fair comparison of the two methods.  
The 2S and 2P pseudopotentials for C are shown in Fig.~\ref{fig1}(a), and we find that the 2S potentials of the two methods differ only very little. 
The differences are more pronounced for the 2P pseudopotentials, in particular 
near $r=0$ where $\mc{V}_1^{\text{TM}}$ and $\mc{V}_1^{\text{PA}}$ differ by about $1\,\text{Ry}$. Next we discuss the pseudopotentials of the TM and PA methods for $\text{Fe}^{2+}$  
shown in Fig.~\ref{fig1}(b). We find that the 3S and 3P potentials are very similar for both methods. On the contrary, the 3D potential of the PA 
and  TM methods differ by $\approx 5\,\text{Ry}$ near $r=0$. Furthermore, the TM potential for the 3D state ``oscillates'' around the corresponding PA potential, 
i.e., the potentials cross twice near $r\approx 0.6a_0$ and $r\approx 0.9 a_0$. This behaviour 
is consistent with the fact that the degree of the polynomial describing $\mc{V}_l^{\text{TM}}$ in the core region is more than 
twice as large as in the case of~$\mc{V}_l^{\text{PA}}$.

\begin{figure}[!t]
\begin{center}
\includegraphics[width=1\columnwidth]{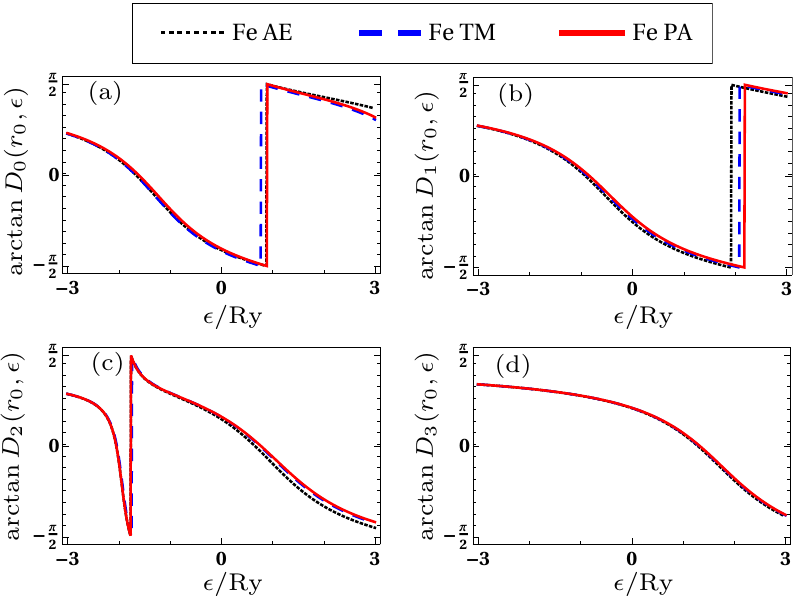}
\caption{Arctan of the logarithmic derivative defined in Eq.~(\ref{logder}) for Fe${}^{2+}$ at $r_0=2.4 a_0$ as a 
function of energy $\epsilon$. The results of the AE, TM and PA calculations are shown by black dotted, blue dashed and red lines, respectively. The 
angular momentum channels shown are (a) $l=0$, (b) $l=1$, (c) $l=2$ and (d) $l=3$. }
\label{fig2}
\end{center}
\end{figure}

In the case of iron we performed tests of pseudopotential transferability
for different oxidation states and different occupations of the 3d orbitals. 
In Tab.~\ref{tab:transf} we report the energy difference between the AE and pseudopotential 
eigenvalues  for a set of spin-polarised configurations. 
We find that the energy differences for the TM and PA pseudopotentials are  of the same order of magnitude, 
and no clear trend distinguishing the two methods can be established. It follows that 
the TM and PA pseudopotentials are both transferable to a similar extent. 
\begin{table}[!t]
\begin{center}
\begin{tabular}{l|cc|cc}
\hline\hline
& \multicolumn{2}{c|}{Fe TM} & \multicolumn{2}{c}{Fe PA} \\
& \multicolumn{2}{c|}{$\Delta E/\text{Ry}$} &  \multicolumn{2}{c}{$\Delta E/\text{Ry}$} \\
Configuration & $\mathrm{3d}\uparrow$  & $\mathrm{3d}\downarrow$   & $\mathrm{3d}\uparrow$   & $\mathrm{3d}\downarrow$  \\
\hline
Fe$^{0\phantom{+}}$: 3d$^5_1$ 4s$^1_1$ & \mm0.0166 & $-$0.0072 & \mm0.0223 & $-$0.0016 \\
Fe$^{0\phantom{+}}$: 3d$^4_2$ 4s$^1_1$ & \mm0.0132 & \mm0.0012 & \mm0.0187 & \mm0.0067 \\
Fe$^{2+}$: 3d$^5_1$ 4s$^0_0$           & \mm0.0083 & $-$0.0166 & \mm0.0086 & $-$0.0161 \\
Fe$^{2+}$: 3d$^4_2$ 4s$^1_0$           & \mm0.0081 & $-$0.0132 & \mm0.0117 & $-$0.0087 \\
\hline\hline
\end{tabular}
\caption{Energy difference $\Delta E$  between AE  and pseudo-eigenvalues  for different
spin-polarised configurations of the iron atom.}
\label{tab:transf}
\end{center}
\end{table}
In order to investigate the transferability of the pseudopotentials generated via the TM and PA methods further, we introduce the 
dimensionless logarithmic derivative
\begin{align}
D_l(r_0,\epsilon) = \left. a_0 \frac{\text{d}}{\text{d}r}\log R_l(r,\epsilon)]\right|_{r=r_0}\,,
 \label{logder}
\end{align}
where $a_0$ is the Bohr radius and  $R_l$ is the radial wavefunction corresponding to  energy  $\epsilon$ and orbital angular momentum $l$. 
$R_l$ can be either an AE wavefunction or a PWF, and for an ideal pseudopotential the logarithmic derivatives of the PWFs and the AE wavefunctions 
match over a wide range of energies. A comparison of $D_l(r_0,\epsilon)$ for AE, TM and PA wavefunctions in the case of Fe${}^{2+}$ is shown in 
Fig.~\ref{fig2} for four different angular momentum channels. We find that the logarithmic derivatives of the TM and PA wavefunctions are very similar, 
and both of them follow the results of the AE calculation closely. This result is consistent with our findings in Tab.~\ref{tab:transf} and shows that 
the PA  method results in pseudopotentials that are as transferable as their TM counterparts.

\begin{figure}[!t]
\begin{center}
\includegraphics[width=1\columnwidth]{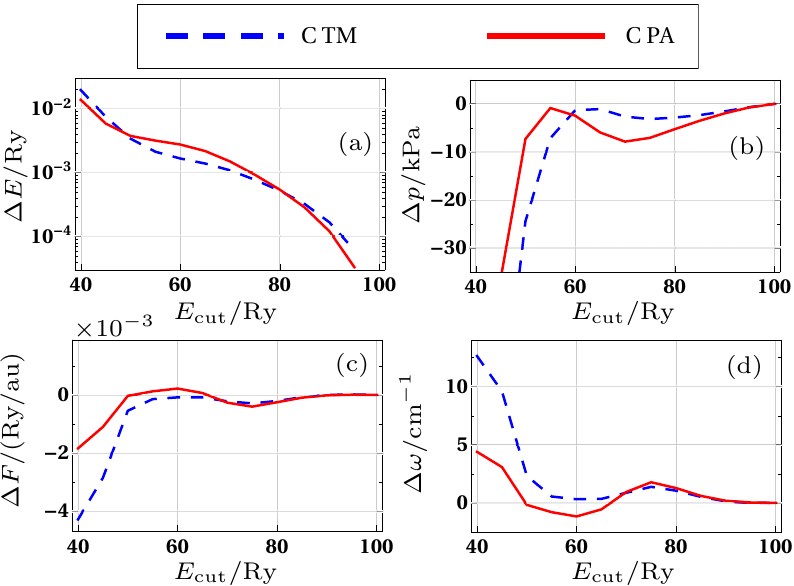}
\end{center}
\caption{Electronic structure calculations for C in the diamond structure. Convergence of (a) total energy $E$, (b) pressure $p$, (c) $x$-component of force $F$ acting on atom 1 and  (d) 
zone-center optical phonon mode frequency $\omega$. In each panel we show 
$\Delta X= X(E_{\text{cut}})- X(E_{\text{cut}}^{\text{max}})$ with $X\in\{E,p,F,\omega\}$ and $E_{\text{cut}}^{\text{max}}=100~\text{Ry}$. Results from the PA (TM) method are 
shown by red solid (blue dashed) lines.}
\label{fig3}
\end{figure}

\begin{figure}[!t]
\begin{center}
\includegraphics[width=1\columnwidth]{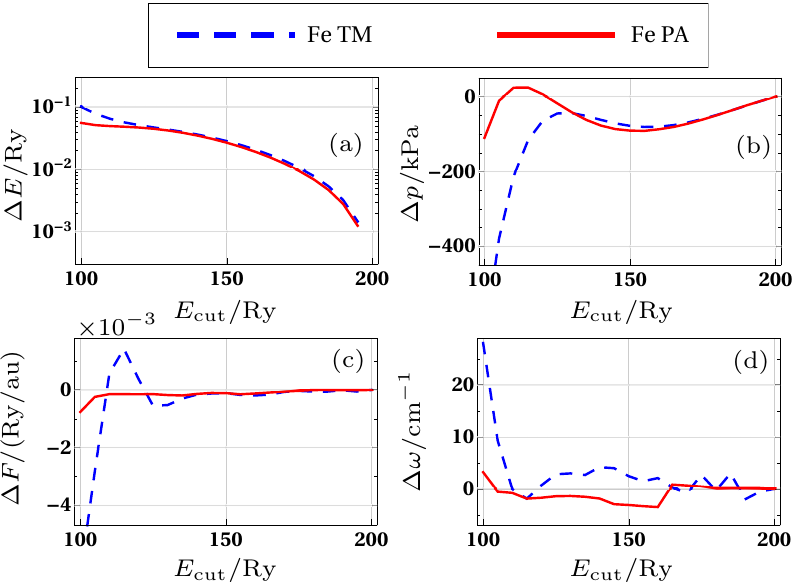}
\end{center}
\caption{Electronic structure calculations for  Fe in the \emph{bcc} structure. Convergence of (a) total energy $E$, (b) pressure $p$, (c) $x$-component of force $F$ acting on atom 1 and (d) 
zone-center optical phonon mode frequency $\omega$.  In each panel we show 
$\Delta X= X(E_{\text{cut}})- X(E_{\text{cut}}^{\text{max}})$ with $X\in\{E,p,F,\omega\}$ and $E_{\text{cut}}^{\text{max}}=200~\text{Ry}$. 
Results from the PA (TM) method are shown by red solid (blue dashed) lines.}
\label{fig4}
\end{figure}

Next we consider bulk electronic structure calculations for carbon and iron. 
In all calculations we use the Kleinman-Bylander representation~\cite{kleinman:82} of the pseudopotentials 
and choose the pseudopotential with the largest angular momentum as the local potential. 
First, we calculate the equilibrium lattice constants  of diamond and \emph{bcc} iron. 
To this end we evaluate the energy vs. volume curve and fit it to the Vinet equation of state~\cite{vinet:87}. 
The results are reported in Tab.~\ref{tab:eos} and show that the two PP methods result in practically equivalent 
results for carbon. In the case  of iron, differences between the two pseudopotential methods are also small. 
\begin{table}[!b]
\begin{center}
\begin{tabular}{llll}
\hline\hline
Atom & $a$ (\AA) & $B_0$ (GPa) & $B^\prime$ \\
\hline
C\, TM & 3.5402 & 458.1 & 3.558 \\
C\, PA & 3.5434 & 456.3 & 3.609 \\
\hline
Fe\, TM & 2.7956 & 189.8 & 6.769 \\
Fe\, PA & 2.8104 & 184.7 & 8.458 \\
Fe\, FP-LAPW & 2.7479 & 193.1 & 4.170 \\
\hline\hline
\end{tabular}
\end{center}
\caption{Calculated equilibrium lattice spacing $a$, bulk modulus $B_0$ and pressure derivative
of bulk modulus $B^\prime$ according to the Vinet equations of state, for diamond and bcc iron.  
All-electron FP-LAPW calculations were performed with the ELK code (v4.3.6)~\cite{elk}
using the PZ functional, smearing and k-point sampling.}
\label{tab:eos}
\end{table}
Results from an AE calculation for iron are shown in the last row of Tab.~\ref{tab:eos} and demonstrate that 
both pseudopotential methods overestimate the equilibrium lattice spacing by about 2\%. 
The  lattice spacing obtained from the TM method is 0.5\% smaller  than the one obtained with the PA method 
and thus slightly closer to the AE result.
Note that the pressure derivative of the bulk modulus $B^\prime$ in Tab.~\ref{tab:eos} takes on large values for all three methods in the case of iron. 
This problem has been reported in~\cite{Zhang2010} and is due to an incipient magnetic transition
at volumes larger than the equilibrium, and is an artefact of the fitting. 
In order to better distinguish  between the two pseudopotential methods, we systematically investigate the convergence of 
the total energy, the pressure and higher derivatives 
of the total energy  with the plane wave cutoff energy $E_{\text{cut}}$.  
For carbon we use the structure of cubic diamond at the experimental
lattice spacing of 3.567~\AA, while for iron we use the ferromagnetic \emph{bcc} structure
with a lattice spacing of 2.870~\AA. 
In order to rule out any possible bias from the 
experimental lattice spacing, we also perform calculations at other values  and obtain
qualitatively identical results.  
The convergence of the total energy and pressure with  E$_\text{cut}$ 
for carbon is shown in Figs.~\ref{fig3}(a) and~(b), respectively. 
We find that the rate of convergence for these two quantities is similar for both methods. 
The corresponding results for iron are shown in Figs.~\ref{fig4}(a) and~(b), and here the 
PA pseudopotential gives rise to a well converged pressure at smaller values of $E_{\text{cut}}$ compared to the TM method. 

The convergence of the first and second derivatives of the total energy
with respect to atomic coordinates is investigated as follows. For analysing the first energy derivative, 
we displace one atom by $1\%$ of the lattice constant in the $x-$direction
and extract the value of the restoring force as a function of $E_{\text{cut}}$. Information about the convergence of the 
second energy derivatives is obtained by calculating the frequency of the zone-center optical phonon mode in the 
un-distorted lattice. 
The results for carbon are shown in Figs.~\ref{fig3}(c) and~(d), and illustrate that the PA potentials give better results 
for forces and phonon frequencies at small cutoff energies $E_{\text{cut}}$  compared to the TM method. 
This trend is even more pronounced for iron as can be seen in Figs.~\ref{fig4}(c) and~(d). In particular, the 
phonon frequency calculation converges much faster for the PA pseudopotentials compared to the TM method.

Finally, for completeness we provide the converged values of the total energy, pressure, force and phonon frequencies for C and Fe in  Tab.~\ref{tab:abs}.  
We find that both pseudopotential methods result in very similar values for most quantities. The largest deviation occurs for pressure in the 
case of Fe, where the  TM and PA values differ by~$\approx 25\%$. This discrepancy is related  to the fact that the TM and PA pseudopotentials 
give rise to slightly different equilibrium lattice constants as shown in Tab.~\ref{tab:eos}.  
On the other hand, the differences between TM and PA values for  total energy, force and phonon frequency are less than $4\%$ for both C and Fe. 
\begin{table}[!b]
\begin{center}
\begin{tabular}{lllll}
\hline\hline
Atom & $E$ (Ry) & $p$ (kPa) & $F$ (Ry/au) & $\omega$ ($\text{cm}^{-1}$)\\
\hline
C\, TM & -22.870 & -101.5 & -0.2356 & 1286.47\\
C\, PA & -22.858 & -89.2 & -0.2362 & 1288.14\\
\hline
Fe\, TM & -245.207 & -182.4 & -0.0365 & 221.04\\
Fe\, PA & -244.487 & -145.0 & -0.0379 & 222.80\\
\hline\hline
\end{tabular}
\end{center}
\caption{Total energy $E$, pressure $p$, force $F$ and phonon frequency $\omega$ for diamond with lattice spacing 3.567~\AA\ 
 and \emph{bcc} iron with lattice spacing 2.870~\AA.}
\label{tab:abs}
\end{table}
\section{Summary and discussion\label{sum}}
In this paper we have introduced a novel method for generating NCPPs. We represent  the pseudopotential by a polynomial of degree ten in the radial variable 
and impose the same set of smoothness conditions as the TM method~\cite{troullier:91}. The latter approach  makes an Ansatz for the PWF and obtains 
the pseudopotential by an inversion of the radial Schr\"odinger equation. The resulting TM potential  is also a polynomial 
in the radial variable, but its degree of twenty-two is significantly higher than the polynomial representing our pseudopotentials. 
A direct Ansatz for the pseudopotential instead of an Ansatz for the wavefunction thus  allows us to impose the 
same set of smoothness conditions as the original TM method, while at the same time 
the degree of the polynomial representing the pseudopotential can be significantly reduced. 
This reduced polynomial degree  comes at an increased numerical cost for the generation of our pseudopotentials compared to the 
TM method. The numerically most expensive step for generating a TM potential is in finding a solution to the norm conservation condition, 
see Eq.~(\ref{nctm}). Each integral requires $\mc{O}(N)$ operations, where $N$ is the number of points in the discrete grid. On the other hand, 
our method requires to solve an eigenvalue problem on the same grid and thus needs $\mc{O}(N^3)$ operations. However, we find that the search 
for the optimal coefficients resulting in  a norm-conserving potential with the correct ground state energy converges quickly with Newton methods.

In Sec.~\ref{results} we compared our PA pseudopotentials with the TM method in electronic structure calculations for carbon and iron in the \emph{bcc} structure. 
A comparison of the screened PA and TM pseudopotentials for C and $\text{Fe}^{2+}$ reveals that the two methods result in similar potentials, and 
the largest differences occur for the states with highest angular momentum. Moreover, we find that the TM potentials of higher angular momentum states 
cross the PA potentials several times as a function of the radial variable. These ''oscillations`` are a signature of the higher-order polynomials in 
the TM potentials. 
The main result of this work is that our PA potentials speed up the convergence of the first and second derivatives of the total energy
with respect to atomic coordinates. More specifically, we find  our new  scheme 
produces accurate forces and Hessians at lower cutoff energies than the TM method. This could constitute a
big computational saving when calculating structural relaxations and phonon frequencies 
in large systems. It follows that the increased cost in generating our PA pseudopotentials is by far outweighed by 
the promised savings in large-scale electronic structure calculations. 
\begin{acknowledgments}
The authors acknowledge financial support from the National Research Foundation
and the Ministry of Education, Singapore. DJ acknowledges funding from the 
European Research Council under the European Unionʼs Seventh Framework Programme 
(FP7/2007-2013)/ERC Grant Agreement no. 319286, Q-MAC. 
\end{acknowledgments}

\appendix

\section{Coefficients \label{coeff}}

\vspace*{-0.2cm}
The expressions for the  coefficients $\mc{C}_{2i}$ in Eq.~(\ref{C_TM}) in terms of $c_{2i}$ are given by 
\begin{subequations}
\begin{align}
 \mc{C}_0 & = \mc{E}_l+(4l+6)c_2 \,, \\
 \mc{C}_2 & = 4\left[(2l+5)c_4 + c_2^2\right] \,, \\
 \mc{C}_4 & = 6(2 l+ 7)c_6 + 16 c_2c_4 \,, \\
 \mc{C}_6 & = 8\left[(2 l+ 9)c_8 + 3 c_2c_6 +2 c_4^2 \right]  \,, \\ 
 \mc{C}_8 & = 10(2 l+ 11)c_{10} + 32 c_2c_8 +48 c_4 c_6 \,, \\
 \mc{C}_{10} & = 4\left[ (6 l+ 39)c_{12} + 10 c_2c_{10}+ 16 c_4 c_8 + 9 c_6^2\right] \,, \\ 
 \mc{C}_{12} & =  16\left[ 3 c_2c_{12}+ 5 c_4 c_{10} + 6 c_6 c_8 \right] \,, \\ 
 \mc{C}_{14} & = 8\left[ 8 c_8^2 + 12 c_4 c_{12} + 15 c_6 c_{10}\right] \,,\\ 
 \mc{C}_{16} & = 16 \left[ 9 c_6 c_{12}+ 10 c_8 c_{10} \right] \,, \\ 
\mc{C}_{18} & =  4\left[25 c_{10}^2 + 48 c_8 c_{12}\right] \,,\\ 
 \mc{C}_{20} & =  240 c_{10} c_{12} \,,\\ 
 \mc{C}_{22} & =  144 c_{12}^2 \,.
\end{align}
 \label{TM_coeff}
\end{subequations}
The coefficients $\mc{X}_{6}$, $\mc{X}_{8}$ and $\mc{X}_{10}$ in Eq.~(\ref{C_PA}) as a function of 
$\mc{X}_{0}$ and $\mc{X}_{4}$ are ($\mc{X}_{2}=0$)
\begin{subequations}
\begin{align}
 \mc{X}_6   = & -\frac{1}{8\rc^6}\left[ 80\mc{X}_0+24\rc^4 \mc{X}_4 
 -80\mc{V}_{\text{AE}}(\rc)\right. \notag \\
  & \left. \qquad \qquad +17 \rc V_{\text{AE}}^{\prime}(\rc)  -\rc^2V_{\text{AE}}^{\prime\prime}(\rc) \right] \,,\\
 \mc{X}_8 = & -\frac{1}{4\rc^8}\left[- 60\mc{X}_0 - 12\rc^4 \mc{X}_4 
 +60V_{\text{AE}}(\rc)\right. \notag \\
  & \left. \qquad \qquad -15 \rc V_{\text{AE}}^{\prime}(\rc)  +\rc^2 V_{\text{AE}}^{\prime\prime}(\rc) \right] \,, \\
 \mc{X}_{10} =& -\frac{1}{8\rc^{10}}\left[ 48\mc{X}_0  + 8 \rc^4 \mc{X}_4 
 -48V_{\text{AE}}(\rc)\right. \notag \\
  & \left. \qquad \qquad +13 \rc V_{\text{AE}}^{\prime}(\rc)  -\rc^2V_{\text{AE}}^{\prime\prime}(\rc) \right] \,.
\end{align}
\label{PA_coeff}
\end{subequations}
%
%
%
%
\end{document}